\documentclass[11pt]{article}
\usepackage[utf8]{inputenc}

\usepackage{graphicx} 
\usepackage[section] {placeins}

\title{Really Small Shoe Boxes -- On Realistic Quantum Resource Estimation}

\author{Alexandru Paler\footnote{Johannes Kepler University, Linz Institute of Technology, Altenbergerstra\ss e 69, 4040 Linz, Austria}, Daniel Herr\footnote{Theoretical Quantum Physics Laboratory, RIKEN Cluster for Pioneering Research, Wako-shi, Saitama 351-0198, Japan and Computational Physics, ETH Zurich, 8093 Zurich, Switzerland} and Simon J. Devitt\footnote{Centre for Quantum Software \& Information (QSI), Faculty of Engineering \& Information Technology, University of Technology Sydney, Sydney, NSW, 2007, Australia and Turing Inc., Berkley, CA, 94705, USA}}

\date{\today}

\date{
}
\begin{document}

\maketitle

\begin{abstract}
    Reliable resource estimation and benchmarking of quantum algorithms is a critical component of the development cycle of viable quantum applications for quantum computers of all sizes. Determining critical resource bottlenecks in algorithms, especially when resource intensive error correction protocols are required, will be crucial to reduce the cost of implementing viable algorithms on actual quantum hardware.  
\end{abstract}

The potential of quantum computing will be proven with quantum computers up and running, but it is uncertain when this will happen. Nevertheless, researchers and companies are racing to build the first large-scale error corrected quantum computers. The private sector seems increasingly optimistic about that the commercial viability of quantum computing will be a breakthrough, but researchers are still somewhat pessimistic about when the first large-scale computer that conclusively solves useful problems better than classical machines will exist. We argue for a pragmatic approach - while it could be sooner rather than later, it is going to take some time before quantum machines begin to be cost effective solutions to a scientifically and/or commercially useful problem. A pragmatic approach to the assessment of quantum applications beings with accurate resource benchmarking -- deriving useful performance analytics as to the actual physical resources consumed by a given quantum computation.  This will is enabled by software, used to estimate the resources consumed by a quantum program, to both assess the current state of quantum hardware and software development and to refine algorithms to speed up their eventual adoption.

\section{The Shoe Box}
Bin packing \cite{Garey:1990:CIG:574848} is a quintessential computer science problem, and very often optimisation problems are equivalent to packing as many objects with different volumes as possible, in a bin of a given volume. We will state right from the beginning that compiling fault-tolerant, error-correcting quantum algorithms is very similar to bin packing, but we call the bins shoe boxes. This is not only done for rhetorical reasons, but because one of the authors used shoe boxes for archiving files related to well defined topics (e.g. hand written computer engineering lecture notes). Each shoe box included the content of a much broader topic, i.e. everything necessary to understand and explain the topic. Once the notes were written (\emph{synthesised}), these were reorganised in the weekends to eliminate redundancy (\emph{optimisation}), learned and reproduced during an exam (\emph{verification}) and finally stored into the box. If the same broad topic needs to be re-learned after many years, the contents of the shoe box is optimal with respect to the learning and re-learning process.
\\
\\
In the following discussion, the difficulty of implementing bin packing for fault-tolerant, error corrected quantum applications (and consequently perform resource optimisation and benchmarking) will be visually explained. There is a significant practical advantage related to this problem: the smaller the shoe box, the sooner that scientifically or commercially viable quantum computing becomes a practical reality. The software tools used to generate shoe boxes are sometimes referred to as quantum compilers, but in later sections we argue that \emph{quantum resources estimators} (the software enabling the realistic view on the quantum computing horizon) are more than compilers and will form in fact the core of future \emph{realistic quantum operating systems} and will become a key component in the developmental cycle of quantum applications.

\subsection{One qubit, two qubits, three qubits, \ldots}

A significant amount of research has been invested into finding out how many qubits a quantum computer will need in order to be more powerful than the most powerful classical computer. This problem is commonly referred to as \emph{quantum supremacy}, where abstract problems are formulated specifically for quantum computers that are classically intractable for classical computers of any size \cite{Boixo:2018aa}. Other formulations of the problem are sometimes known as \emph{quantum advantage} \cite{QAdv}, were other metrics such as dollar costs for running programs on classical and quantum hardware are taken into account. Researchers in both the public and private sector are engaged in active research to define accurate metrics to define when a quantum computing system reaches these regimes. This research is somewhat related to the more formal problem of bin packing \cite{Garey:1990:CIG:574848} -- in this context, what is the largest instance of the most practical\footnote{practical in this context is often related to the ability to use a quantum algorithm to obtain a computational result that is either scientifically or commercially valuable beyond the fact that a quantum computer was used to derive it.} quantum algorithm which can be executed on the largest available quantum computer? In other words, what is the largest quantum computation that can be squeezed to the current (and in the near future) small-sized quantum chips (also known as (N)oisy (I)ntermediate (S)cale (Q)uantum \emph{NISQ} machines \cite{P18}).

The term \emph{small size quantum chip} does not refer to the dimensions of the chip in centimeters, but to some metric that considers both the number of qubits and the topology of the inter-qubit connections. It is not straightforward to define a particularly expressive metric, some use individual 2-qubit gate fidelities as a marker for the overall quality of a qubit array, some use quantum volume \cite{B17} and some place a category boundary between applications that require active quantum error correcting codes and applications that do not. To this date there is no consensus within the community and therefore, it is not exactly clear what \emph{small} refers to exactly. But it is almost certain that the current generation of qubit chips, with less than 100 qubits, are small. With these devices, entering the quantum supremacy or quantum advantage regime may be possible, but it is still very unclear if practical quantum computing, using NISQ machines, are even possible.

\subsection{Multiplied by computing time,}

The connections supported between qubits within a given chip determine the number of operations necessary to execute an algorithm. The problem is very similar to how computations are executed on classical machines: if not all the registers are addressable from a given register, or by a certain instruction, some workarounds have to used. These workarounds increase the time necessary to execute the computation.

One of the main issues is that the quantum hardware is not perfect, such that the information stored in the quantum registers (the physical  hardware qubits) loses its precision (\emph{decohere}) over time. Furthermore, each quantum instruction (known as a \emph{gate}) has a negative effect on precision. Therefore, because hardware (both storage and operation) is faulty, in the absence of error correction \cite{Devitt:2013aa}, a quantum computation has to be executed as fast as possible and with as few gates as possible.  If the NISQ regime is defined as the boundary where resource heavy quantum error correction protocols need to be used, then these computations need to be executed with fast and with few physical qubits.  A reasonably accurate rule of thumb is to take the number of qubits needed for a given algorithm, $Q$, and the {\em depth} of your algorithm, $K$ (depth is the number of parallel gate steps needed in the algorithm) and calculate the quantity, $A = 1/(KQ)$.  If the physical error rates of your qubit array (the worst error rate associated with qubit initialisation, single and two-qubit gates and measurement over the entire computer), $p$, is $p < A$, then your algorithm is potentially implementable without error correction.  If $p \gg A$, then extensive error correction will be needed and your application is highly unlikely to be NISQ compatible.  

\subsection{Equals a shoe box}

The research community is increasingly agreeing on the need for abstracting the computational resources (hardware and time) necessary to execute a particular computation as a \emph{space-time volume}, i.e. the number of physical qubits needed for a particular algorithm (space), multiplied by the gate depth of the algorithm (time) \cite{Paler:2017aa}.  In this discussion, we will call this the shoe box. The volume of a computation depends on many factors, and once more, there is no exhaustive study which enumerates all of them. 

Determining the shoe box, which encapsulates a quantum algorithm, in principle benchmarks how large a quantum computer will be needed to execute it. When this is done rigorously, all practical quantum algorithms are shown to be too large to run on the current generation of physical hardware. Thus, there is a real gap between the available quantum chips/computers and the algorithms one would like have executed. As a result, three general approaches are being taken:
\begin{itemize}
    \item Increase the chip sizes - more qubits, better connectivity.
    \item Improve the algorithms - less hardware and/or less execution time.
    \item Optimisation of shoe boxes - the same algorithm in smaller volume.
\end{itemize}

We argue that in order to efficiently accelerate hardware development using these three broad approaches, we require flexible, and fast software solutions to resource estimate and benchmark quantum algorithms as accurately as possible (not just in terms of final qubit and computational time metrics, but detailed analytics about where resources are used and what are the most costly subcomponents of a quantum algorithm). Thus, we focus the following discussion on this notion of quantum performance analytics and benchmarking and how the shoe box problem can be used to accurately analyse the performance and hardware requirements of a quantum program.

\begin{figure}
	\includegraphics[width=\linewidth]{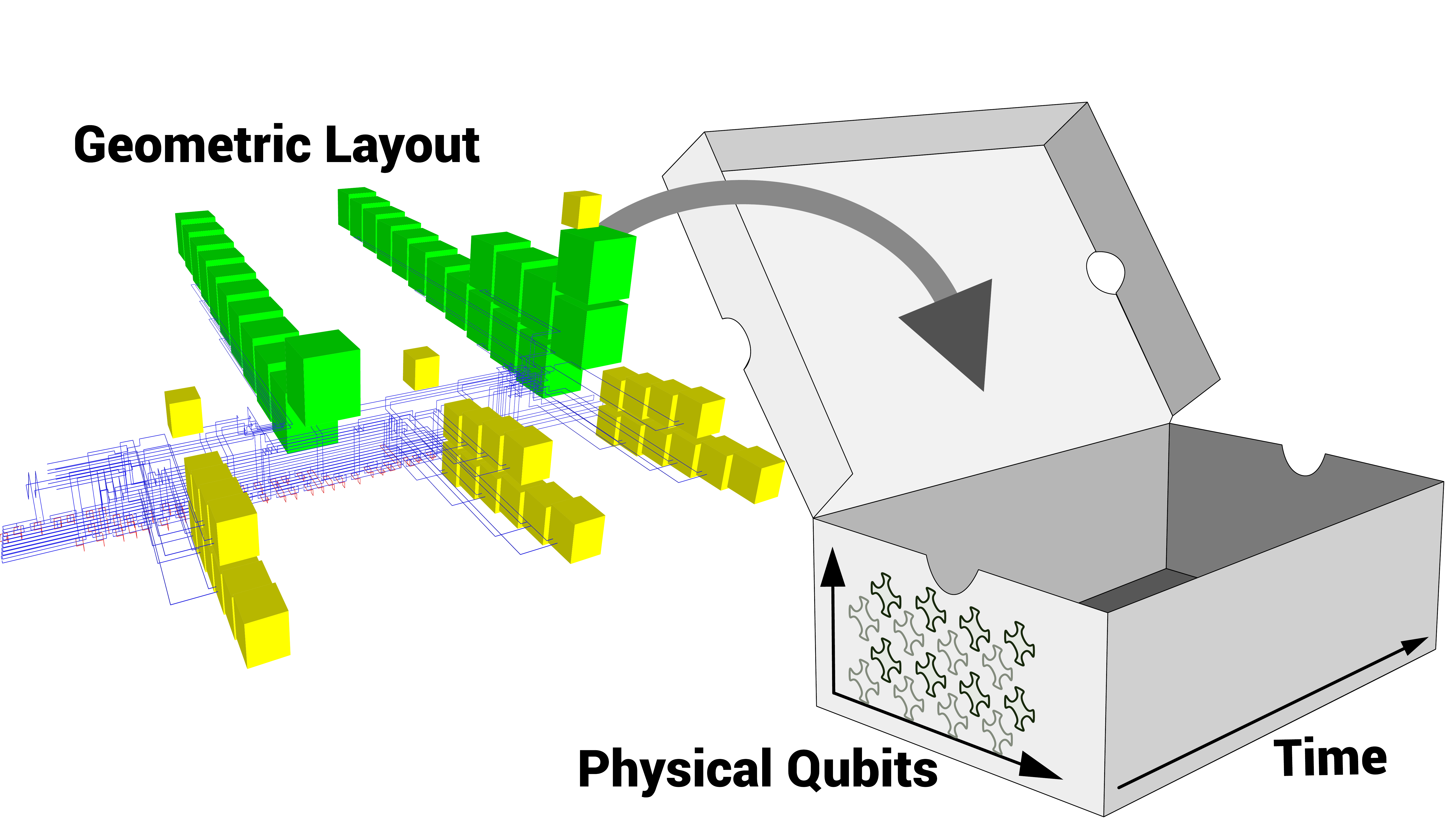}
    \caption{\label{fig:shoe_box} The algorithm on the left needs to fit into the shoe box on the right. The dimensions of the shoe box are given by the number of physical qubits and the computational time of the algorithm.}
\end{figure}

\subsection{Error correction: replacing paper with clay tablets}

The shoe box analogy was introduced in conjunction with lecture notes written on sheets of paper. These lecture notes stand in relation to the environment they are in (i.e. they are coupled to a bath of entropy). This motivates another analogy: a quantum computer is like a the bath tub filled with water, and executing a quantum algorithm is to drop the shoe box in the bath tub. Although paper is light, easy to write on, and cheap to produce, the problem is that it can be easily destroyed by the water. A solution would be to use clay tablets (properly baked) instead of paper. This comes with advantages and disadvantages: the contents of the shoe box would be relatively immune to damage from the water, but the volume of the box would be much larger because each clay tablet is now a big, bulky and robust object, that nevertheless contains no more information than the original, delicate piece of paper. 

However, there does not seem to be another solution in practice. Quantum hardware is so fragile that error-correction has been accepted as a necessity for applications of ``practical" size. Therefore, there appears little hope for the current NISQ machines\footnote{If the definition of NISQ is chosen to be applications that do not need error correction.} to be able to execute industrially relevant quantum computations. Such machines have too few qubits to support error correction for the entire quantum program. Thus, as with clay tablets, optimisation of shoe boxes includes the following two research directions:
\begin{itemize}
    \item Smaller clay tables for the same information content -- minimise the number of qubits necessary for error detection and correction
    \item Thinner clay tablets with same strength -- minimise the time necessary for performing error correction, without reducing its resilience to errors.,
\end{itemize}

\section{Technical Background}

Quantum resource estimation is the process of synthesising (compiling), optimising and verifying an error-corrected quantum computation which requires the least physical resources in terms of qubit hardware and execution time -- think of writing on the thinnest and smallest possible clay tables. Realistic quantum resource estimation requires us to explicitly compile a fully error corrected algorithm (to the level of essentially generating an execution sequence for every qubit and every gate) and to map this to a given hardware model using the most up to date technical assumptions. The vast majority of algorithmic benchmarking is done on a case-by-case basis, often by hand, and necessitates significant hardware and compilation assumptions \cite{PhysRevX.8.041015}. As is often the case, conservative  assumptions are chosen. These assumptions are too conservative for some researchers, while for others not conservative enough. A lot of assumptions also relate to design issues at the hardware level.  For example, resource benchmarking can occur under the assumptions that:
\begin{itemize}
    \item The theoretically best error correcting code is not the easiest one to implement in hardware.
    \item Hardware capabilities are not as advanced as error correction would require.
\end{itemize}

We argue, that for the moment, every assumption should be as conservative as possible (pessimism). The goal is to start from large, thick clay tablets and invent plastic sheets on the way (newer techniques in error corrected compilation and optimization), by making as few technical assumptions as possible and building a platform that can adapt to changes in experimental hardware design on the fly. In the following we present the technical aspects which determine the size and the number of the clay tablets in a shoe box, which can then be used to benchmark and analyze a fully error corrected quantum algorithm.

\subsection{Surface Codes}

Quantum error correction is a component of all practical quantum computing architectures, and a wide variety of codes have been proposed. However, one of the codes most utilised, in practice, is the surface code, because it can tolerate high error rates (compared to other codes) while having one of the most straightforward required hardware configurations (also, compared to other codes). Consequently, surface codes are very often used as architectural building blocks.

Error correction can be implemented with different techniques using the surface code \cite{FMMC12}, but all the recipes start from the same ingredients: 1) construct an array of qubits arranged in a two-dimensional lattice; 2) allow each qubit to interact with all its direct neighbours (e.g. for qubits not on the boundary: north, south, east, west) and be able to perform qubit specific initialisation and measurement in multiple bases. Two of the most common techniques are called \emph{braiding} and \emph{lattice surgery}. The latter has recently attracted more attention, because it seems to require less physical qubits to implement an identical computation \cite{FG18}. Braiding starts from the assumption that holes  (also called \emph{defects}) can be punctured into the imaginary surface spanned across the qubit lattice. Quantum algorithms are encoded into the movement of the holes around the surface. Surgery uses a different concept: the spanned surface is split into \emph{patches}, which can be glued together (merged) or  cut apart (split). For this technique, quantum algorithms are encoded in the sequence of logic operations that are enacted on encoded patches when they are merged and split.

\begin{figure}
	\includegraphics[width=\linewidth]{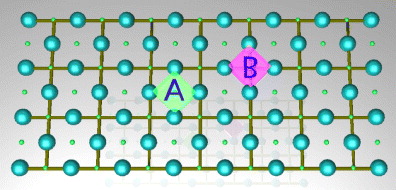}
    \caption{\label{fig:surface_code} This is a planar patch of surface code. All large circles represent data qubits which encode a single logical qubit. The small circles (two types marked A and B) are ancillary qubits that continuously measure the parity of the surrounding data qubits, and extract information related to physical qubit errors.}
\end{figure}

Similar to classical error correction, quantum codes have a distance -- the number of errors which can be detected and corrected. For the surface code, the distance is related to the physical ($x-y$) dimensions of the holes (when braiding) or of the patches (when performing surgery). The distance is chosen based on an error model that captures the hardware characteristics which are considered relevant. Faulty hardware will require strong error correction, thus longer code distances, which means more qubits. Increasing the quality of the qubits reduces the amount of necessary error correction.  This amounts to hardware optimisation in the benchmarking formalism and needs to be built into any analytics platform as an automatic update when higher fidelity operations are reported for any potential hardware system.

The number of qubits used for error correction could be also optimised when keeping the hardware quality constant, by arranging the holes or the patches more efficiently on the surface -- a case of two-dimensional bin packing \cite{herr2017optimization}. The time necessary to execute a surface code protected quantum computation can be reduced by determining shorter hole movements or shorter sequences of patch interactions.  This is the software optimisation in the benchmarking problem.

An accurate and useful quantum resource estimator assumes that hardware will reach the minimum necessary accuracy for error-correction to work.  This necessitates software optimisation as the key component to a reliable analytics platform and that shoe box volumes can be optimised by:
\begin{itemize}
    \item more compact encoding of the algorithm on an imaginary two-dimensional surface of qubits -- reduces lattice dimensions $\rightarrow$ less qubits
    \item more compact, but computationally equivalent, interaction of the holes or patches on the surface -- less steps necessary to move a hole $\rightarrow$ less execution time on the computer $\rightarrow$ less exposure to decoherence (shoe box in the water filled bath tub) $\rightarrow$ shorter distance $\rightarrow$ less qubits.
\end{itemize}

\subsection{Clifford and T(-Bone)}

One of the difficulties of error correction is that arbitrary operations cannot be enacted on encoded data in a simple way (technically known as transversal logic operations). Some  gates in a universal gate set have to be reformulated (rewritten/compiled) into a fault-tolerant form compatible with the code. In practice, this means that, with some code exceptions, computations are decomposed into a sequence of operations (instructions/gates) chosen from a set called Clifford + $T$. This is because the are known recipes of how to encode fault-tolerant qubit states and logic gates from the Clifford + $T$ universal set.

While drafting this manuscript we discovered that ``Clifford the big red dog" is a series of children books \cite{Cliff} that has strangely encapsulated the qualitative nature of this problem. The cartoon series includes another dog called T-Bone. Clifford is friendly and helpful, and T-Bone is a bulldog with a large appetite. The character traits of the cartoon dogs are surprisingly consistent with the effect of the Clifford and $T$ gates on surface code protected computations.

Computations consisting entirely of Clifford gates are friendly, because these can be efficiently simulated on classical computers. Thus, universal quantum computations require something more than only Clifford gates. This extra component is realised by the $T$ gate, where $T^4 = Z$,  $Z$ is the Pauli phase flip gate and Clifford + $T$ computations are universal and cannot be efficiently simulated on classical computers. When implementing error corrected computations, at least one gate is difficult to implement directly in the code space, and for surface codes this the $T$ gate. Each time a $T$ gate has to be applied in a fault-tolerant, error corrected manner, an additional complex mechanism is included into the computation: a \emph{distillation} procedure \cite{FD12}. These procedures, as their name suggest, are used to distill (purify) faulty computational resources into less faulty ones. $T$ gates have a large appetite for computational resources, because each distillation uses additional qubits and time, ultimately to enact a single logic gate.

Researchers have targeted this problem heavily, as reducing the number of $T$ gates in a quantum algorithm results automatically in shoe box volume reduction. At the same time, reducing the volume of distillation protocols to enact individual $T$ gates results in vast improvements, because (at least for the moment) all distillation procedures implement the same gate. Surprisingly, there are computations for which the number of $T$ gates does not dominate the  error correction resource overhead \cite{PhysRevX.8.041015}. The number and properties of such algorithms has not been investigated thoroughly, but it has suggested that in some important cases, Clifford optimisation could be the dominant factor in generating smaller shoe boxes.

\begin{figure}
	\includegraphics[width=\linewidth]{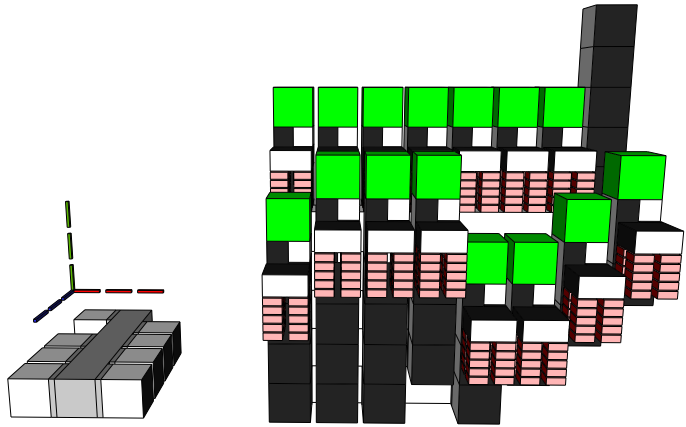}
    \caption{\label{fig:overheadCliffordvsT} A space-time pictorial representation of a quantum algorithm.  Each ``box" contains braiding or lattice surgery patters that enable certain computations in an error corrected space (in this case, time runs vertically and the horizontal cross-section represents the qubit resource requirements for the hardware).  The left half of the figure implements a Clifford operations. The right hand side is the necessary preparation of a single $T$-gate, using distillation and injection (red blocks). The blue and red dashed axes represent the hardware coordinates, and the green axis is time (runs vertically, bottom-up). The figure was first presented in \cite{FG18}}.
\end{figure}

\section{Realistic assumptions}

The previous section noted that assumptions related to hardware, compilation and implementation are, for the moment, generally pessimistic and conservative. The surface code choice and fact that computations are required to be Clifford + $T$, implies that some technical assumptions have already been made. These assumptions may prove to be true as hardware development continues and even with this choices fixed, we still have significant flexibility in regard to how shoe boxes are generated and look: thin (few qubits) and long (long computational times) vs. thick (lots of qubits) and short (fast)? or we may choose more futuristic materials than clay (different QEC codes?), which may require more flexible hardware architectures? Other design considerations include issues such as what is the speed of the fastest classical computer which is economically viable to control the error corrected quantum computer? How does this influence design choices of both the quantum code and the hardware it is run on?

\subsection{One distillation at a time}

Parallelisation of quantum algorithms will only be possible with cheap and scalable hardware. This will not be the case in the foreseeable future as qubits will be expensive and in short supply.  Consequently, in the near term for compilation and benchmarking, it seems that the most sensible decision is to allow distillations to be executed only sequentially: this saves significant amounts of physical hardware. The trade-off is that computations will take longer to execute. It is not possible to speed computations up by being optimistic about future qubit numbers which are not reflected in the available developmental roadmapa for large-scale hardware.

This assumption opens novel research directions, because it has always been assumed that computation depth can be reduced by parallelising $T$ gates. Moreover, depth is routinely reduced by inserting additional qubits into the computation (used as temporary workbenches -- similar to how one adds RAM to a computer when the current processor is too slow, and a new processor too expensive). For the error corrected regime, and in the situation where all qubits require the same error correction, such an optimisation method is not always feasible.

Future work should address new methods to reduce circuit depth without increasing the number of computation qubits. Such approaches would need to account for distillation sequentially, because $T$ gates cannot be executed at any time, but only when their corresponding distillation protocols are finished \cite{paler2018controlling}. It suffices to mention that, for the moment, a distillation circuit is the execution of multiple error corrected Clifford gates, which implies that in the worst case, two sequential error corrected $T$ gates will be separated in time by a block of gates used to perform distillation (i.e. temporal penalties in sequential $T$ gate distillation can be reduced by compacting the space-time volume of individual $T$ gate distillation protocols).

\subsection{The practical alternative to surface codes is the surface code}

There a many alternatives to the surface code, and most of them were proposed because the surface code requires distillation protocols for universality. The hope is that there exist codes which have the same, or even better, error correction properties, and which do not require the inclusion of distillations to achieve universal, error corrected computation. Such codes have been found, but in devil is always in the details.  Many of these codes are not realistic from a hardware engineering point of view. For example, requiring dense, long range qubit connections that continue to expand as the computer scales. That is fine in principle, but unacceptable to any experimental hardware engineer that needs to build large arrays of high fidelity qubits and gates.

The architectural requirements of the surface code are to the advantage of one of the most advanced hardware platform for quantum computers: superconducting qubits. Even so, currently connecting the qubits to support the simplest surface code is a very complex task, and this was visible when the first IBM quantum chips had missing connections between qubits \cite{Corcoles:2015aa}. Increasing the degree of the connections while maintaining and/or increasing their fidelity will be more complex. Thus, the realistic (pessimistic) assumption is that the ability to produce arbitrary, long range connections, with high-fidelity and a maximal amount of parelisability will not be available in the near term.

Most current hardware systems have issues related to scalability.  While qubit chipsets of the order of 1000 physical qubits should be possible, expanding to millions or billions of qubits for fully fault-tolerant, error corrected algorithms will require significant work.  Superconductors suffer from issues related to the cooling needed for large arrays of physical qubits and the ability to send very large numbers of control signals in and out of dilution refrigeration systems.  Ion traps have issues related to creating large vacuum systems to house a large quantum computing system and the intrinsic slow speed of their quantum gates (which become compounded when error correction is implemented).  These scalability issues exist already for surface code based architectures, where qubit manipulation and connectivity is the simplest.  Moving to other coding techniques simply makes these issues worse.

Another significant issue is related to the minimal error rates needed for the codes to become functional (known as the fault-tolerant threshold).  The surface code has one of the highest threshold of any coding technique (around 0.7\%).  More complex codes, that have better asymptotic behaviour or allow for more resource friendly implementation of universal gate sets, also have lower fault-tolerant thresholds.  3D gauge fixed colour codes, which allow transverse gates for the full Clifford + $T$ set, have thresholds simulated without explicit circuit constructions (phenomenological model) of approximately 0.3\% \cite{Brown:2016aa}.  It is expected that these codes would have a full circuit threshold of around 0.05\%, which is more than an order of magnitude below the surface code.  This leads to an interesting question.  Even though gauge fixed colour codes have transversality available for a full universal gate set, error rates needed by hardware will have to be at least 0.01\% (likely lower).  Even if this were possible for the hardware in the short to near term, you would now be sitting nearly two orders of magnitude below the threshold of the surface code.  Consequently, the distance of the surface code for a given quantum circuit would be much lower than the colour code.  It is still unclear if the additional qubit/time resources needed for state distillation at lower distance surface code would be better or worse than not requiring distillation, but needing a larger code distance overall for gauged fixed colour codes?

While many QEC codes have been proposed over the years \cite{lidar2013quantum} (and are still proposed), the experimental constraints of potential hardware systems puts heavy restrictions on what schemes are ultimately practical in the foreseeable future.  It would be interesting for future research to devise and analyse the \emph{emulation} of other codes on surface code architectures. To the best of our knowledge, code emulation has not been looked at, and there are compelling arguments against it. Nevertheless, considering the current state of the art quantum computer proposals and the difficulty of constructing quantum chips, we argue that emulation may be the only feasible option to using different codes in practice. Thus, alternative codes will have to use the surface code architecture under the hood.

\begin{figure}
	\includegraphics[width=\linewidth]{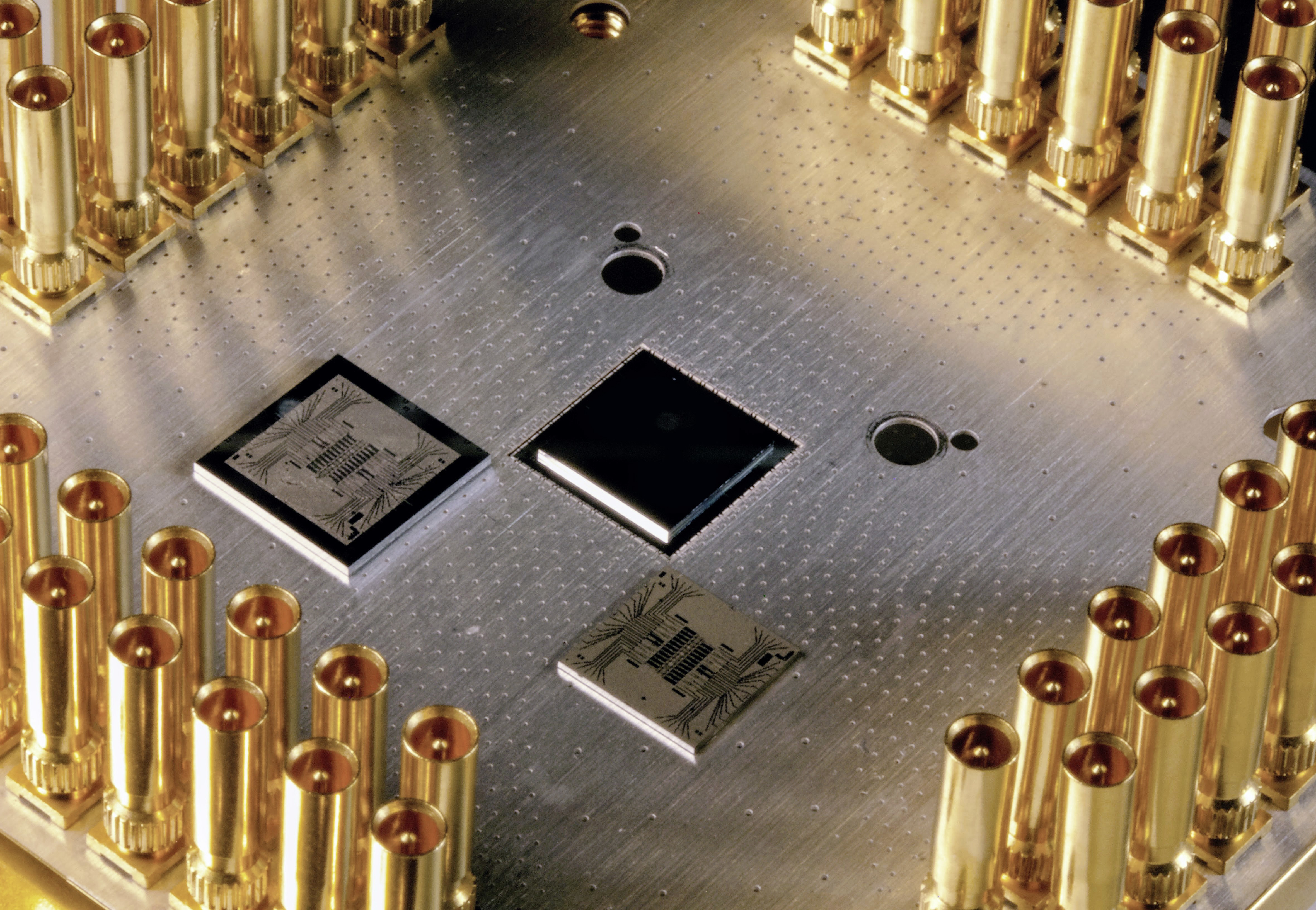}
    \caption{\label{fig:quantum_chip} A quantum chip.}
\end{figure}

\section{Perspectives}

We started discussing shoe boxes, cartoon dogs and continued presenting technical assumptions in order to achieve realistic quantum resource estimations. We return our discussion to the shoe box analogy and argue that its contents is more than just a stack of clay tablets. When packing a quantum computation into a shoe box we implicitly presumed that the computation is static (not dynamic). This means that all the tablets in the box are going to be executed in a well defined and known order. However, in practice there are situations when some tablets will not be executed. This is because the shoe box represents a \emph{worst case estimation} in which, for example, at each branching instruction the handling of each branch is explicitly inserted into the computation: the concept of functions, code reuse etc. is not applied. The shoe box represents the worst case execution of a computation. This is why distillation procedures are repeated throughout the error corrected computation.

The software tool that we described is better to be called resource estimator instead of compiler. Compilers are more advanced resource estimators, because the first are able to identify redundancies in the shoe box, and output a more compact representation of the same box (without affecting its associated volume). The shoe box can be considered a loop unrolled version of the compiled computation. The following two sections discuss the forms in which realistic resource estimators could evolve into.

\subsection{Just in time compiler}

The classical machine controlling the execution of the quantum computer is connected to a permanent feedback loop: signals received from the quantum hardware are interpreted into error correction syndromes, but these could also be used to dynamically adapt the computation. For example, if a distillation procedure failed it could be restarted instead of aborting the entire computation. However, restarting a distillation implies that the contents of the shoe box is dynamically updated $\rightarrow$ the show box is generated just in time (JIT). Moreover, the software tool can be transformed into a JIT compiler that predicts future execution paths in order to reduce shoe box volumes, and which compactifies the representation of the shoe box.

Quantum JIT compilers are still a technological speculation, but their existence could bring quantum computing closer. Similar to how classical JIT compilers integrated components of a runtime system, it seems a realistic design decision to consider that JIT resource estimation/compilation is performed in an hardware abstraction layer managed by a reliable classical runtime system. Such a system  integrates the error correction mechanism too, and is able to dynamically decide code distances of executed components, for example. This will trigger dynamic recompilation (re-placement) of the error corrected structures (holes and patches for surface codes) on the actual quantum hardware.

JIT compilation is a highly complex problem. Thus, it is realistic to assume that the first, commercially or scientifically viable quantum computer will be computationally universal, but able to execute only a specific algorithm (because it is resource optimised for this problem rather than because of constraints on its capability). Fine tuning all the parameters of the error correction and the classical control mechanisms would be much easier if the computer would implement a single type of algorithm (e.g. energy optimisation of a particular molecule) and do it in very resource efficient manner. As a result, only very specific portions of the computation would be JIT. The first versions of the computer, may very well not include any JIT compilation.

\begin{figure}
	\includegraphics[width=\linewidth]{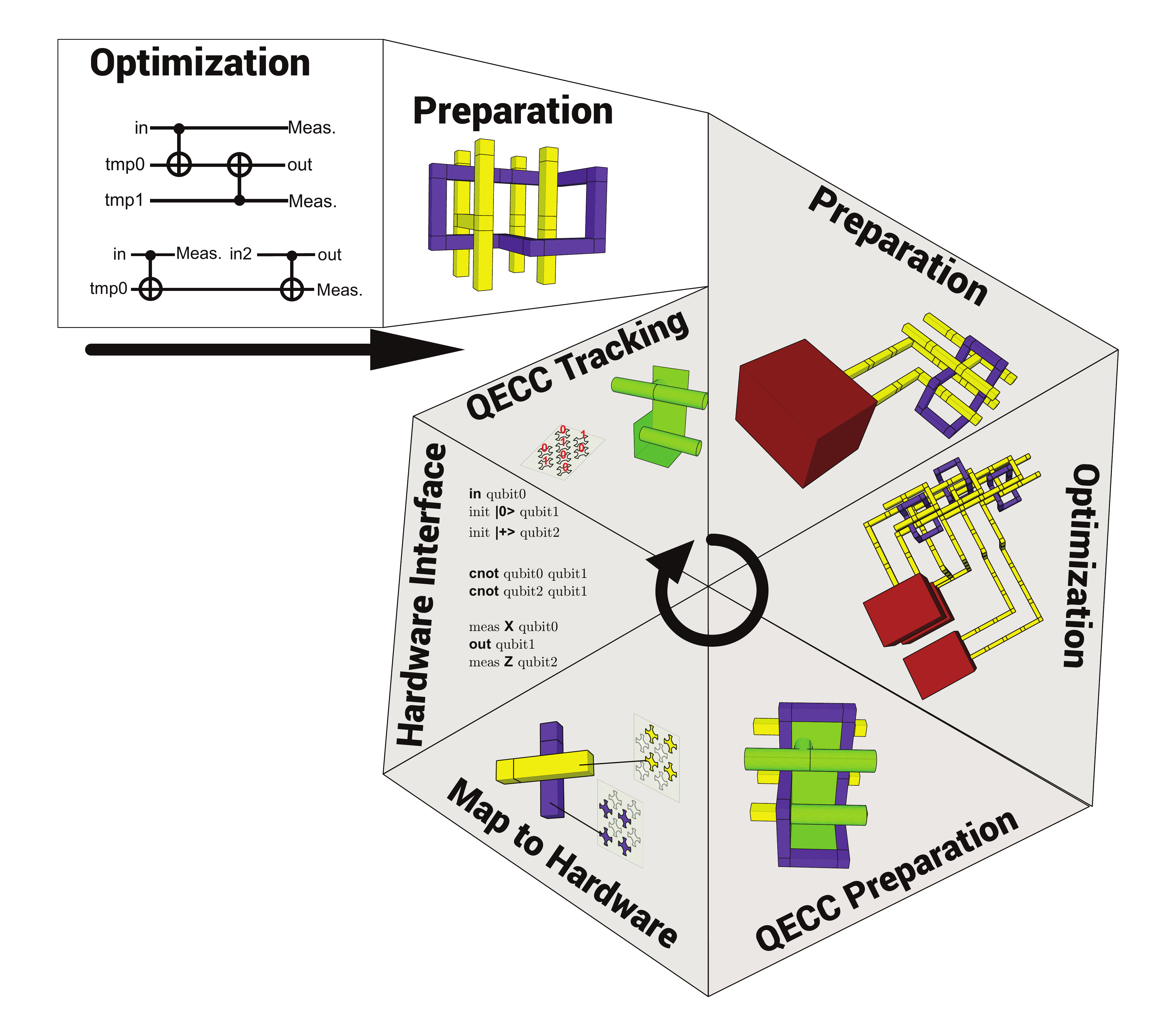}
    \caption{\label{fig:control_software} First is the compilation then error tracking and response to the measurement results. The grey components are executed in a loop, because fault-tolerant circuits have a dynamic structure. The green tubes, red boxes and yellow bars are visualisations of simple braided structures.}
\end{figure}

\subsection{Operating systems}

The reliability of a fault-tolerant large scale quantum computer will be as good as the reliability of its weakest component. It would be a mistake to assume that once quantum computations are fault-tolerantly executed, the entire computer is fault-tolerant. The reliability of the classical computer controlling the quantum machine needs to be taken into consideration. One would not like the quantum operating system equivalent of the blue screens of death when executing an algorithm. 

If JIT compilers will become reality, these will be running inside a runtime system (consider it a straightforward OS) to control quantum hardware. The system ensures that it schedules hardware access in such a way that processes (compilation, error correction) does not reach live locks or other dangerous situations. Therefore, it may actually be argued that JIT compiler execution will be managed by a scheduler. Nevertheless, for now these are technical speculations, which should be taken into consideration when discussing about realistic and practical quantum operating systems.

\section{Perspective}

We refrain from making forecasts and have already mentioned some future work. A realistic resource estimator should be tailored for surface codes and assume that distillations are sequential. The estimator should have both academic and industrial applications, while being very scalable in the sense that it can estimate circuits up to tens of thousands of qubits. Reaching such a development milestone will start a new research and development race for the construction of a classical software control framework for reliable quantum computers. The time horizon for practical quantum computing is uncertain, but it is almost certain that practical resource estimators and systems to analyse the performance of theoretically designed quantum algorithms will be up and running before the first actual commercially or scientifically viable quantum computer.

\bibliographystyle{unsrt}
\bibliography{main}

\begin{thebibliography}{10}

\bibitem{Garey:1990:CIG:574848}
Michael~R. Garey and David~S. Johnson.
\newblock {\em Computers and Intractability; A Guide to the Theory of
  NP-Completeness}.
\newblock W. H. Freeman \&amp; Co., New York, NY, USA, 1990.

\bibitem{Boixo:2018aa}
Sergio Boixo, Sergei~V. Isakov, Vadim~N. Smelyanskiy, Ryan Babbush, Nan Ding,
  Zhang Jiang, Michael~J. Bremner, John~M. Martinis, and Hartmut Neven.
\newblock Characterizing quantum supremacy in near-term devices.
\newblock {\em Nature Physics}, 14(6):595--600, 2018.

\bibitem{QAdv}
Rigetti Computing.
\newblock {Quantum advantage prize rules and criteria},
  https://medium.com/rigetti/quantum-advantage-prize-rules-and-criteria-81c35256e4df
  (2018).

\bibitem{P18}
J.~Preskill.
\newblock {Quantum Computing in the NISQ era and beyond}.
\newblock {\em Quantum}, 2:79, 2018.

\bibitem{B17}
L.S. Bishop, S.~Bravyi, A.W. Cross, J.M. Gambetta, and J.A. Smolin.
\newblock {Quantum volume}, https://dal.objectstorage.open.softlayer.
  com/v1/AUTH\_039c3bf6e6e54d76b8e66152e2f87877/community-documents/quatnum-volumehp08co1vbo0cc8fr.pdf
  (2018).

\bibitem{Devitt:2013aa}
S.J. Devitt, W.J. Munro, and K.~Nemoto.
\newblock Quantum error correction for beginners.
\newblock {\em Reports on Progress in Physics}, 76(7):076001, 2013.

\bibitem{Paler:2017aa}
A.~Paler, I.~Polian, K.~Nemoto, and S.J. Devitt.
\newblock Fault-tolerant, high-level quantum circuits: form, compilation and
  description.
\newblock {\em Quantum Science and Technology}, 2(2):025003, 2017.

\bibitem{PhysRevX.8.041015}
Ryan Babbush, Craig Gidney, Dominic~W. Berry, Nathan Wiebe, Jarrod McClean,
  Alexandru Paler, Austin Fowler, and Hartmut Neven.
\newblock Encoding electronic spectra in quantum circuits with linear t
  complexity.
\newblock {\em Phys. Rev. X}, 8:041015, Oct 2018.

\bibitem{FMMC12}
A.G. Fowler, M.~Mariantoni, J.M. Martinis, and A.N. Cleland.
\newblock {Surface codes: Towards practical large-scale quantum computation}.
\newblock {\em Phys. Rev. A.}, 86:032324, 2012.

\bibitem{FG18}
A.G. Fowler and C.~Gidney.
\newblock {Low overhead quantum computation using lattice surgery}.
\newblock {\em arXiv:1808.06709}, 2018.

\bibitem{herr2017optimization}
Daniel Herr, Franco Nori, and Simon~J Devitt.
\newblock Optimization of lattice surgery is np-hard.
\newblock {\em npj Quantum Information}, 3(1):35, 2017.

\bibitem{Cliff}
Norman Bridwell.
\newblock {Clifford the Big Red Dog},
  https://en.wikipedia.org/wiki/Clifford\_the\_Big\_Red\_Dog (2018).

\bibitem{FD12}
A.G. Fowler and S.J. Devitt.
\newblock {A bridge to lower overhead quantum computation}.
\newblock {\em arXiv:1209.0510}, 2012.

\bibitem{paler2018controlling}
Alexandru Paler.
\newblock Controlling distilleries in fault-tolerant quantum circuits: problem
  statement and analysis towards a solution.
\newblock In {\em Nanoscale Architectures (NANOARCH), 2018 IEEE/ACM
  International Symposium on}. IEEE, 2018.

\bibitem{Corcoles:2015aa}
A.~D. C{\'o}rcoles, Easwar Magesan, Srikanth~J. Srinivasan, Andrew~W. Cross,
  M.~Steffen, Jay~M. Gambetta, and Jerry~M. Chow.
\newblock Demonstration of a quantum error detection code using a square
  lattice of four superconducting qubits.
\newblock {\em Nature Communications}, 6:6979 EP --, 04 2015.

\bibitem{Brown:2016aa}
Benjamin~J. Brown, Naomi~H. Nickerson, and Dan~E. Browne.
\newblock Fault-tolerant error correction with the gauge color code.
\newblock {\em Nature Communications}, 7:12302 EP --, 07 2016.

\bibitem{lidar2013quantum}
Daniel~A Lidar and Todd~A Brun.
\newblock {\em Quantum error correction}.
\newblock Cambridge University Press, 2013.

\end{thebibliography}

\end{document}